\documentclass[11pt,twoside]{article}

\usepackage{graphicx}
\usepackage{asp2006}
\usepackage{epsf}
\usepackage{psfig}
\usepackage{lscape}

\begin{document}
\title{Outflow and Accretion in Massive Star Forming Regions}   
\author{P.D. Klaassen \& C.D. Wilson}   
\affil{McMaster University, 1280 Main St. W., Hamilton ON, Canada \mbox{L8S 4M1}}    

\begin{abstract} 
In order to distinguish between the various components of massive star forming regions (i.e. infalling, outflowing and rotating gas structures) within our own Galaxy, we require high angular resolution observations which are sensitive to structures on all size scales. To this end, we present observations of the molecular and ionized gas towards massive star forming regions at 230 GHz from the SMA (with zero spacing from the JCMT) and at 22 and 23 GHz from the VLA at arcsecond or better resolution.  These observations (of sources such as NGC7538, W51e2 and K3-50A) form an integral part of a multi-resolution study of the molecular and ionized gas dynamics of massive star forming regions (i.e. Klaassen \& Wilson 2007).  Through comparison of these observations with 3D radiative transfer models, we hope to be able to distinguish between various modes of massive star formation, such as ionized or halted accretion (i.e Keto 2003 or Klaassen et al. 2006 respectively).
\end{abstract}

\section{Introduction}

At the large distances to massive star forming regions, we do not yet quite have the resolution necessary to determine whether or not they can form via a scaled up version of the processes responsible for the formation of low mass stars.  The energetics (i.e. luminosity, outflow energy and accretion rates) are orders of magnitude higher than those seen in low mass star forming regions (e.g. Arce et al. 2007), and the outward radiation and thermal pressures produced when the massive star begins ionizing its surroundings are enormous.  Yet, collimated outflows and both ionized and molecular infall signatures are present in high mass star forming regions just as they are in their lower mass counterparts.

If we assume that massive stars can form in a single accretion event similar to their low mass counterparts, how then can accretion continue beyond the formation of the H{\sc II} region? Does accretion continue in an ionized form? Does it stop early in the star's evolution? Does it continue through a disk (not yet observed around forming O stars)?

G10.62-0.38 (hereafter G10.6) is a well studied ultracompact H{\sc II} (UCH{\sc II}) region (Ho et al. 1981, Keto et al. 1988, etc) through which the surrounding gas is falling in, becoming ionized, and continuing to fall in towards the central star.  The high resolution observations of Keto \& Wood (2006) in H66$\alpha$ taken at the VLA show an infall signature in their PV diagram, and so it appears as though accretion onto the protostar is, in this case, ionized and is continuing beyond the formation of the UCH{\sc II} region.

G5.89-0.39 (hereafter G5.89) is the site of an UCH{\sc II} region (the archetypal shell UHCII region from Wood \& Churchwell 1989) and multiple outflows depending on the molecular tracer observed. This region appears to be the formation site for a number of massive stars including the O5V star which is believed to power the H{\sc II} region (Feldt et al. 2003) and the 1.3 mm continuum source believed to be powering the SiO outflow detected by Sollins et al. (2004).  Recent maser emission studies of this region (Stark et al. 2007, Fish et al. 2005) also suggest that the two sources are independent. The star identified in the mid-infrared by Feldt et al. (2003), corresponding to `G5.89 center' in Stark et al (2007), appears to be less embedded than the Sollins et al. (2004) 1.3 mm source, suggesting it is possibly more evolved than the 1.3 mm source which drives the accelerating SiO outflow.  To date, there are no published CO maps of this region at high enough resolution to determine which of these two sources (separated by $\sim2''$) lies at the center of the large scale decelerating outflow. However, the velocity gradient in the 1667 MHz maser emission towards G5.89 center (Stark et al. 2007) is consistent with the direction of the large scale east-west CO outflow observed by Klaassen et al. (2006) and Watson et al. (2007).

Klaassen et al. (2006) suggested that accretion onto the source powering the large scale outflow may have halted at the onset of the H{\sc II} region. This conclusion appears consistent with the O5 star being responsible for the H{\sc II} region, `G5.89 center' maser emission and the large scale CO outflow,  while the 1.3 mm continuum source from Sollins et al. (2004) is a younger protostar responsible for the SiO outflow.

These studies of G10.6 and G5.89 suggest that high neutral accretion rates (up to 10$^{-2}$ M$_{\odot}$/yr, Edgar \& Clarke 2003) and continued ionized accretion can both account for the formation of massive stars. Yet, these conclusions are based on the in depth studies of two individual regions, and from these two regions, we cannot make conclusive statements about the general nature of massive star formation.  In order to determine whether there is a preferred method, we need to observe these processes in a number of regions, and build up statistics about whether/how accretion onto a massive protostar occurs beyond the onset of dynamical expansion in the UCH{\sc II} regions.

\section{Initial Survey}

We undertook single pointing observations of 23 massive star forming regions at the James Clerk Maxwell Telescope (JCMT\footnote{The James Clerk Maxwell Telescope is operated by The Joint Astronomy Centre on behalf of the Science and Technology Facilities Council of the United Kingdom, the Netherlands Organisation for Scientific Research, and the National Research Council of Canada.}) in SiO (J=8-7), H$^{13}$CO$^+$ and HCO$^+$ (J=4-3).  The sources for this survey were selected based on the presence of UCH{\sc II} regions (from Wood \& Churchwell 1989, and Kurtz et al. 1994), and previously observed outflowing gas  in CS (from Plume et al. 1992, Shirley et al. 2003).  Additional sources which also contained H{\sc II} regions and outflows were taken from Hunter (1997).

Of the 23 sources in this survey, 14 were detected in SiO, which suggests that the outflows in these regions are still moving outward and shocking the ambient gas.  The non-detections of SiO are unlikely to be due to beam dilution since the average distance to the sources which were not detected in SiO is smaller than the distances to the sources which were detected in SiO. Since SiO depletes out of the gas phase approximately 10$^4$ yr after a shock passes through the region (Schilke et al. 2001), we concluded that the outflow sources without SiO detections were older.

Ten of the sources in our survey had double peaked line profiles in HCO$^+$. Of these 10 sources, nine had blue peaks which were brighter than their red peaks.  This over abundance of brighter blue shifted line peaks is suggestive of infall, provided that the H$^{13}$CO$^+$ line, which was shown to be optically thin, has a single peak, and that peak occurs at the same line of sight velocity as the absorption feature in the HCO$^+$ (i.e. Gregersen et al. 1997, Fuller et al. 2005). Of these nine sources with brighter blue peaks, eight had HCO$^+$ profiles with only one peak.  Thus, we determined that we have detected large scale molecular infall towards eight of our 23 sources.

Of the eight sources with infall signatures, seven correspond to sources in which we detected SiO.  This suggests that  half of the sources with active outflows have ongoing infall (Klaassen \& Wilson, 2007).  If we infer that where there is large scale infall, there is small scale accretion, this suggests that half of the sources which have outflows are actively accreting. From our molecular data alone, we cannot tell whether accretion is ionized or neutral. However, these results suggest that both continued (ionized?) and halted accretion are possible formation mechanisms for more massive star forming regions than just G10.6 and G5.89.

\section{High Resolution $^{12}$CO Observations of Massive Star Forming Regions}

In order to determine the outflowing gas dynamics in massive star forming regions, we need to map the molecular outflows at much higher resolutions than those possible with a single dish telescope.

We observed five massive star forming regions at 230 GHz using the SMA in its extended configuration, at a resolution of approximately 1$''$. The 2 GHz passband of the SMA receivers allowed us to detect many lines, including $^{12}$CO (J=2-1), which we observed at a spectral resolution of 0.53 km s$^{-1}$ (see Klaassen et al. 2008 for a detailed description of the observations).

Because $^{12}$CO is the most abundant molecule observable in the conditions typical of a giant molecular cloud, it is generally seen in large scale structures. Since the SMA is an interferometer, it is only sensitive to structures on scales between the longest and shortest baselines of the interferometer configuration. In its extended configuration, and at 230 GHz, the SMA is sensitive to structures on size scales of approximately 1$''$ to 10$''$.  Structures which are larger than this (approximate) upper limit are filtered out, and are missing in the final image.

\begin{figure}[hb!]
\begin{center}
\includegraphics[scale=0.5,angle=-90]{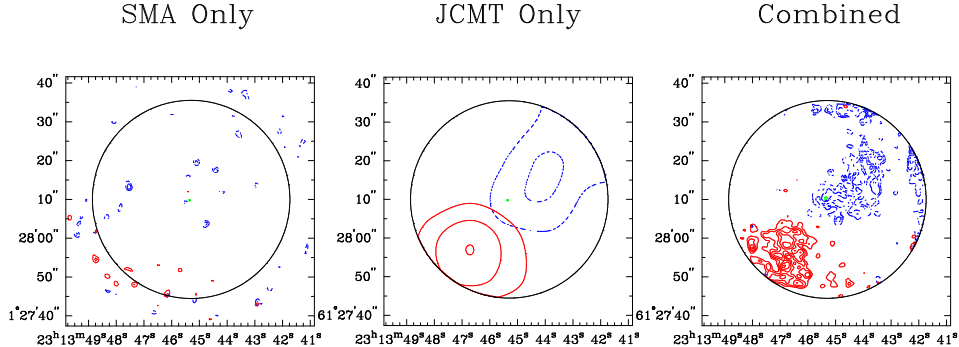}
\caption{Interferometer (SMA) and single dish (JCMT) observations of $^{12}$CO towards NGC7538 IRS 1. Using a non-linear image combination technique (the MOSMEM command in MIRIAD), we were able to combine the data from the two telescopes, resulting in the panel on the right.  The contours for the blue (-74 to -65 km s$^{-1}$, dashed contours) and red (-45 to -34 km s$^{-1}$, solid contours) start at 6$\sigma$, and increase in intervals of 3$\sigma$ for that map.  The green contours in the center of each image represent the 230 GHz continuum emission from IRS 1. The large circle in each panel represents the 45\% gain limits of the SMA at 230 GHz. (Color Figure)}
\label{fig:ngc7538}
\end{center}
\end{figure}

The so called `zero spacing' flux can be observed with a single dish telescope, and then combined with the interferometer data to recover the large scale structures within the primary beam of the interferometer. In the case of the SMA, the JCMT is ideally suited to fill in the center of the UV plane. The outflows from our five sources were barely detectable ($\sim 7\sigma$ detection for NGC 7538) with the SMA alone. However, after combining the SMA and JCMT datasets (using the MIRIAD command MOSMEM), the outflows were clearly observed ($>20\sigma$ in NGC 7538). We were definitely missing flux in the interferometer only dataset (See Figure \ref{fig:ngc7538}). The outflows from massive stars are themselves massive, energetic and large.

From our combined maps, we were able to determine the outflowing mass, momentum, and energy using zeroth and first moment maps. Once we determined the highest velocities of the gas in the outflows, and their lengths, we were also able to determine approximate kinematic ages for these sources, and determine both the outflow luminosity and mass loss rate in the flow (see Klaassen et al. 2008). Preliminary results on the dynamics of the outflowing gas are presented in Table \ref{tab:SMA_kin}.

\begin{table}
\begin{center}
\begin{tabular}{lcccccc}
\hline
\hline\\
&G10.6&G20.08&G28.20&NGC7538&\multicolumn{2}{c}{W51e2}\\
&&&&&LV&HV\\
\multicolumn{7}{c}{Mass (M$_{\odot}$)}                  \\ \hline
Blue    &42.3   & 0.03& 6.8       &10.8   &483&3.3\\
Red     &36.5   & 0.02& 1.6      &2.9    &1057&3.81\\
Total   &1187   & 0.08& 192     &501    &7207&5103\\\hline
\multicolumn{7}{c}{Momentum (M$_{\odot}$ km s$^{-1}$)}\\ \hline
Blue    &204.9  & 0.18& 3.5      &226    &1291&26.4\\
Red     &99.2   & 0.14& 14.4     &113    &2624&33.9\\
\hline
\multicolumn{7}{c}{Energy $\times 10^{42}$ (J)}\\ \hline
Blue    &45.5   & 0.02& 0.07     &47.9   &215&26.2\\
Red     &12.3   & 0.03& 5.4     &44.5   &407&37.5\\
 \hline
\multicolumn{7}{c}{Luminosity $\times 10^{3}$ (L$_{\odot}$)}\\ \hline
Blue    &66     & 0.03& 0.4        &94.5   &146&17.8\\
Red     &18     & 0.04& 32.4     &87.7   &276&25.5\\
\hline
\multicolumn{7}{c}{Mass Loss  Rate $\times 10^{-5}$ (M$_{\odot}$ yr$^{-1}$)} \\ \hline
Blue    &74             & 0.05& 49.1      &25.7   &396&2.7\\
Red     &64             & 0.03& 11.6      &6.9    &864&3.1\\
\hline
\end{tabular}
\caption{Outflow dynamics from data combined from the SMA and JCMT.}
\label{tab:SMA_kin}
\end{center}
\end{table}

If we determine an average distance to massive star forming regions from a number of surveys (i.e. Plume et al. 1992, Shirley et al. 2003, Thompson et al. 2006 and Klaassen \& Wilson 2007) to be approximately 5.6 kpc, the arcsecond resolution at which we observed our sources gives us a spatial resolution of about 5600 AU (or, approximately 0.03 pc).  Observations of disks around massive protostars have suggested disk diameters of order 10000 AU (i.e. Beltran et al. 2006), suggesting we are getting towards the scales on which possible accretion disks could begin to be resolvable.

However, with only two beams across a possible accretion disk, we cannot determine the kinematics of the gas on these small scales, and whether accretion is occurring through these disks. In order to determine whether our observations are resolving critical length scales, we must go to higher resolution and see if we can draw the same conclusions about the small scale structures.

\section{High Resolution VLA observations of K3-50A}

\begin{figure}[bt!]
\begin{center}
\includegraphics[scale=0.6,angle=-90]{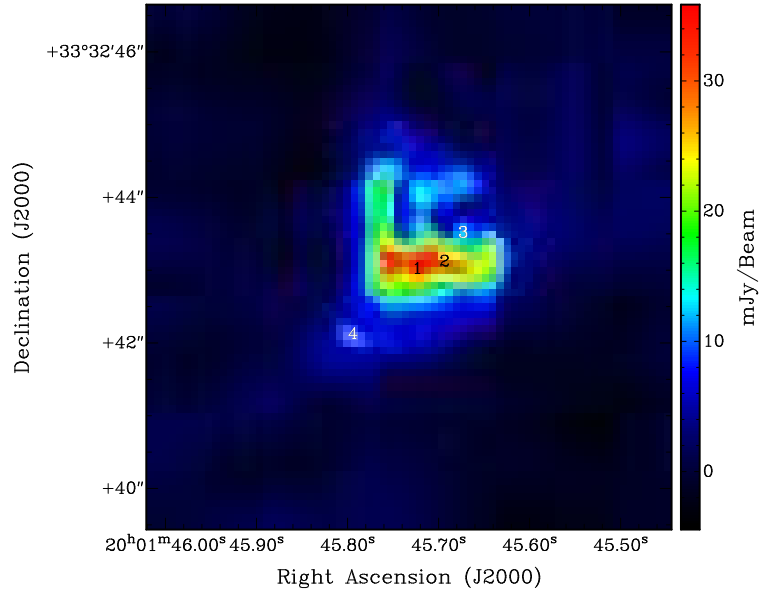}
\caption{22 GHz continuum image towards K3-50A.  The ionized bipolar outflow previously observed by De. Pree et al. (1994) extends to the north and south of the continuum structure seen here.  The four labeled sources are the individual massive protostars identified by Okamoto et al. (2003) in the mid infrared. (Color Figure)}
\label{fig:K3-50}
\end{center}
\end{figure}

Despite being at a large distance (9 kpc), the individual massive protostars in K3-50A have been resolved by mid-infrared observations at a resolution of 0.4$''$ (3600 AU, Okamoto et al. 2003).  We have observed this source at the VLA in its B configuration, obtaining a slightly better resolution of $\sim 0.3''$.  Our 22 GHz continuum image (Figure \ref{fig:K3-50}) is consistent with the findings of Okamoto et al. It appears as though we are observing the free-free emission surrounding the most massive stars (OKYM 3 and 4) in this region. The $\sim$ 18,000 AU flattened structure seen in Figure \ref{fig:K3-50} is perpendicular to the ionized outflow seen in H76$\alpha$ by De Pree et al. (1993), and in ammonia (E. Keto, private communication).  As yet, we have not determined the kinematics of the gas in this flattened structure.

\section{Future Work}

It appears that massive stars can form via accretion despite the outward radiation and thermal pressures presented by the ionizing radiation from the central star.  This accretion may extend beyond the initial dynamical expansion phase of the UCH{\sc II} region (i.e. G10.6), however how much further is uncertain (i.e. G5.89).

 Given a high enough dynamic range in the observations, we can link the small scale structures (observed with interferometers) to the larger scale structures (observed with a combination of interferometers and single dish telescopes and by single dish telescopes alone) and constrain how mass is flowing in these regions.

We are currently running 3D radiative transfer models of protostellar outflows in order to determine what kind of mechanisms can be powering these outflows.  We have modified E. Keto's radiative transfer code to include a number of parameterized outflow generating models.  These models include disk winds (Ouyed \& Pudritz 1997), Parker winds (Keto et al. 1991), bow shocks (Hatchell et al. 1999), photoevaporative winds (Hollenbach et al. 1994) and stellar winds (Mendoza et al. 2004).

We will simulate observing the results of these models with the SMA and JCMT (using the same image combination technique described above for the observed outflows) in order to be able to compare the results of the models to our observations.  Through this comparison, we should be able to constrain what type of mechanism(s) is (are) responsible for the outflows from these massive protostellar regions.  Once we constrain how the outflows are being generated, we can use this as a constraint on the formation mechanism for massive stars.

Because we have observed a large number of sources at resolutions ranging from 15$''$ to 0.3$''$, we should be able to determine which spatial scales {\it need} to be resolved in order to differentiate between outflow generation models, and how they relate to massive star formation.  With this knowledge, we will be able to take full advantage of the resolutions attainable with ALMA.

\acknowledgements 

We would like to acknowledge the support of the National Science and Engineering Research Council of Canada (NSERC).


\end{document}